\documentclass[twocolumn,longbibliography,secnumarabic,amssymb, nobibnotes, aps, pra, showkeys]{revtex4-2}
\usepackage[utf8]{inputenc}
\usepackage[version=4]{mhchem}

\usepackage{amsmath, amssymb, nicefrac, mathtools, makecell, siunitx, empheq, subfig} % Required for inserting images
\usepackage{physics}
\AtBeginDocument{\RenewCommandCopy\qty\SI}
\graphicspath{{figures}}

\newcommand{\PMA}{Division of Physics, Math and Astronomy, California Institute of Technology, Pasadena, California 91125, USA}
\DeclareMathSymbol{\mhyphen}{\mathord}{AMSa}{"39}
\DeclareSIUnit{\ecm}{e \mhyphen cm}
\newcommand{\He}{\ce{^3He}}

\begin{document}

\title{Demonstration of magic dressing of \He}
\author{Raymond Tat}
\affiliation{\PMA}

\begin{abstract}
  A common concern in high-precision neutron electric dipole moment (nEDM) experiments is that of magnetic field stability. For static fields, this problem can be mitigated through the use of a superconducting holding field coil, which when operated in persistent current mode serves to stabilize the magnetic field. However, such a solution is not viable when time-varying magnetic fields are present, as is the case for the spin dressing mode of the proposed nEDMSF (nEDM super-fluid) experiment. This experiment features an oscillating magnetic field to dress the gyromagnetic ratios of \He{} and neutrons to the same value, a condition known as critical dressing. Fluctuations in the dressing field amplitude have the potential to disrupt this condition. Here, we investigate a modification to spin dressing, termed ``magic dressing,'' which renders the system insensitive to small variations in dressing field amplitude. We further demonstrate the utility of this method for the single spin-species case using a sample of polarized \He{}. We find a dramatic increase in transverse relaxation time in the presence of magnetic field gradients.
\end{abstract}

\maketitle

\section{Spin Dressing and nEDM Measurement}
The nEDMSF experiment \cite{nEDMSNS} proposes to measure the neutron electric dipole moment (nEDM) by measuring the Larmor frequency of neutrons under magnetic and electric fields. In such an experiment the addition of a comagnetometer species, which precesses in the same volume as the neutrons, is often included to correct for variations in magnetic field over time. For nEDMSF, this species is \He{}, whose precession frequency is determined by an array of SQUID gradiometers. Monitoring the spin-dependent capture rate of the reaction $n + \He{} \rightarrow \ce{^3H} + p + \qty{765}{\kilo\electronvolt}$ allows for a measurement of the difference between the \He{} and neutron Larmor frequencies. nEDMSF has a secondary mode of operation, termed ``critical dressing mode,'' in which an off-resonant oscillating magnetic field is applied perpendicular to the static magnetic field. This has the effect of modifying the gyromagnetic ratios of the \He{} and neutrons, with the modification factor being given by a Bessel function whose argument is proportional to the magnetic field amplitude. Specifically, for a static field of magnitude $B_0$ and an oscillating field of amplitude $B_1$ and frequency $\omega$, the dressed Larmor frequencies are given by
\begin{align}
    \omega_n &= \gamma_n B_0 J_0(\gamma_n B_1/\omega) \\
    \omega_3 &= \gamma_3 B_0 J_0(\gamma_3 B_1/\omega),
\end{align}
where $\gamma_n$ and $\gamma_3$ are the (undressed) gyromagnetic ratios of the neutron and \He{}, respectively. The argument of the Bessel function, $\gamma B_1/\omega$, is known as the dressing parameter. With an appropriate choice of $B_1$ and $\omega$, one can have both species precess at the same rate. This condition, termed ``critical dressing,'', has a variety of advantages for nEDM measurement. These advantages were analyzed in detailed by Golub and Lamoreaux \cite{GOLUB19941}. However, applying a dressing field comes with its own challenges - whereas the static magnetic field $B_0$ can be fixed through the use of a superconducting coil, this is not possible for the dressing field, whose amplitude must be kept constant over time to maintain the critical dressing condition. Furthermore, any gradients in the dressing field can lead to spin relaxation, resulting in loss of polarization and increased measurement uncertainty. Therefore it is important to consider ways to mitigate variations in the dressing field. Several such strategies were considered by the authors of \cite{QuantumControlPaper}, including using feedback on the dressing amplitude and taking the signal difference between the two measurement cells of nEDMSF. Here we investigate another approach, where we modify the dressing parameters so that the system as a whole is insensitive to changes in the dressing field amplitude, a technique we refer to as ``magic dressing.''

\section{Magic Dressing}
\label{sec:magic-theory}
The concept of a ``magic'' operating condition, wherein a system is rendered insensitive to variations in a parameter, is not new. For example, the ``magic wavelength'' \cite{magicwavelength} can be used to mitigate the AC Stark shift for atoms in an optical lattice. We investigate this idea for use in an nEDM experiment, where the goal is to tune the amplitude and frequency of the dressing field to minimize decoherence. We first examine how this can be done for a single spin species, then explore how this can be extended to the critical dressing mode for nEDMSF.
\subsection{Magic Dressing for a Single Species}
For a single species (i.e. neutrons or \ce{^3He} alone), we can set the dressing amplitude and frequency such that the dressing parameter $x_3 = \gamma_3 B_1/\omega$ is near a local extremum of the zeroth-order Bessel function. In other words, we want
\begin{equation}
\label{eq:1}
    \dv{\omega_3}{B_1} = 0,
\end{equation}
where $\omega_3 = \gamma_3 B_0 J_0(x_3)$. By operating near an extremum of $J_0$ (or equivalently, a zero of $J_1$), the precession frequency can be made insensitive to small changes in $x_3$. The single-species magic dressing technique works both in the case where the dressing field amplitude varies in space, as is the case for inhomogeneities caused by inaccuracies in dressing coil construction, or time as could be caused by fluctuations from the power supply controlling the current. This technique was studied using the spin-exchange optical pumping (SEOP) apparatus discussed in section \ref{sec:SEOP}. As UCN were not available at the time of writing, only single-species magic dressing is investigated experimentally here.

\section{SEOP Apparatus}
\label{sec:SEOP}
We now describe the apparatus used to produce polarized \ce{^3He} at Duke. These subsystems include the laser optics, the $B_0$ system, the dressing field ($B_1$) system, the NMR electronics, and the SEOP cell itself.

\subsection{Laser Optics}
\begin{figure}
\centering
\includegraphics[width=0.7\linewidth]{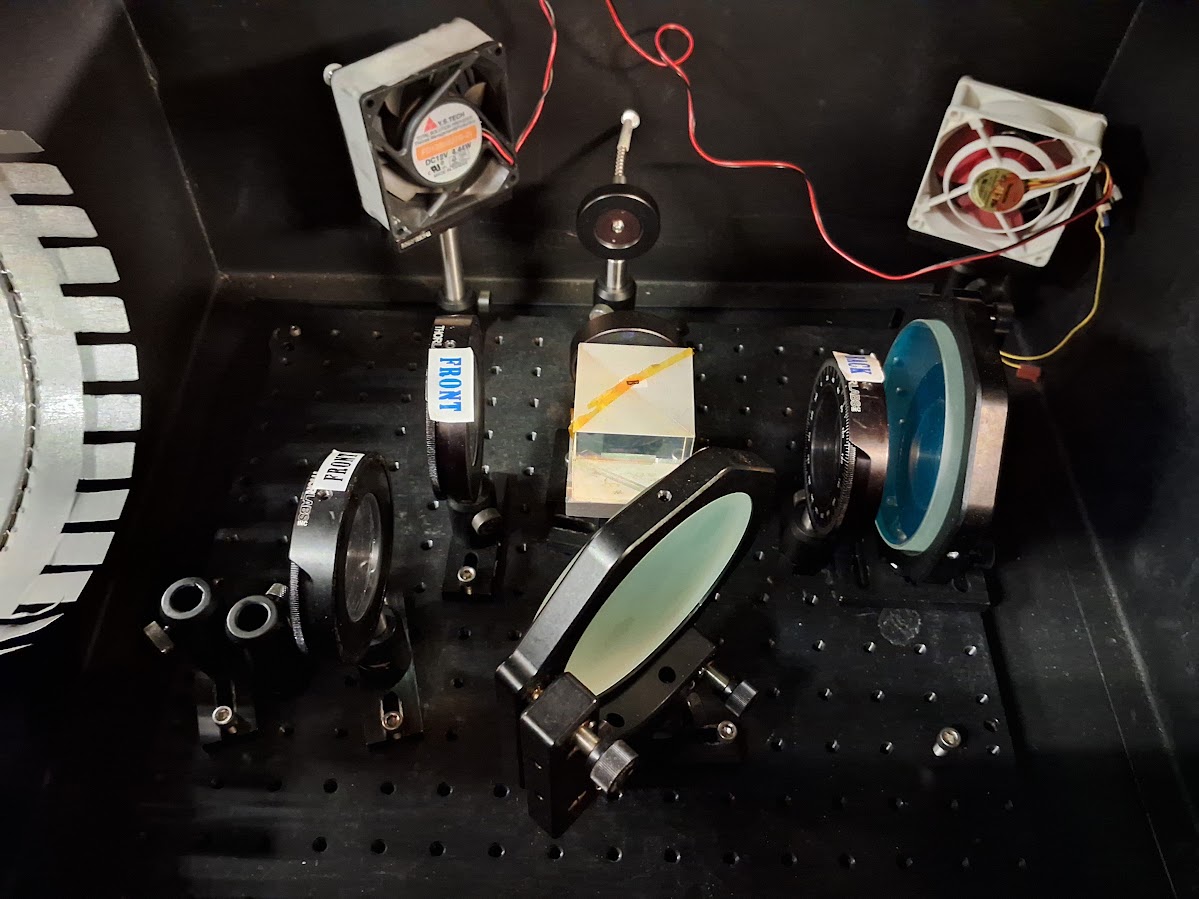}
\caption{Optics elements to produce two beams of circularly polarized light to optically pump the rubidium vapor.}
\label{fig:optics_pic}
\end{figure}

\begin{figure}
\centering
\includegraphics[width=0.7\linewidth]{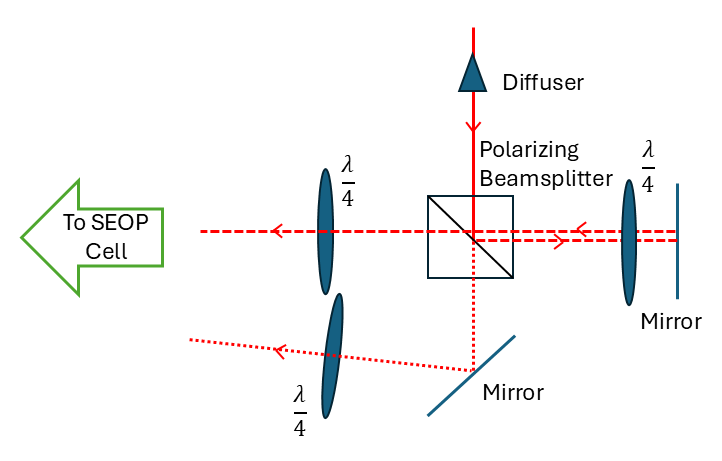}
\caption{Schematic of optics used for the SEOP system. Having two beam paths that converge at the cell allows us to utilize all \SI{60}{\watt} of laser power.}
\label{fig:optics_diagram}
\end{figure}

The optics used to produce circularly polarized \SI{795}{\nano\meter} laser light is shown in figure \ref{fig:optics_pic} and is represented schematically in figure \ref{fig:optics_diagram}. The \SI{60}{\watt} beam first hits a beam diffuser (which ensures that the beam is spread across the entire cell). The polarizing beamsplitter cube separates the two orthogonal components of polarization. The vertically polarized light passes through a quarter wave plate and is subsequently circularly polarized. The horizontally polarized beam first passes through a quarter wave plate twice (thus making it vertically polarized), and then passes through a quarter wave plate to make it circularly polarized with the same handedness as the other beam.

\subsection{Magnetic Field System}
\begin{figure}
    \centering
    \includegraphics[width=0.7\linewidth, angle=-90]{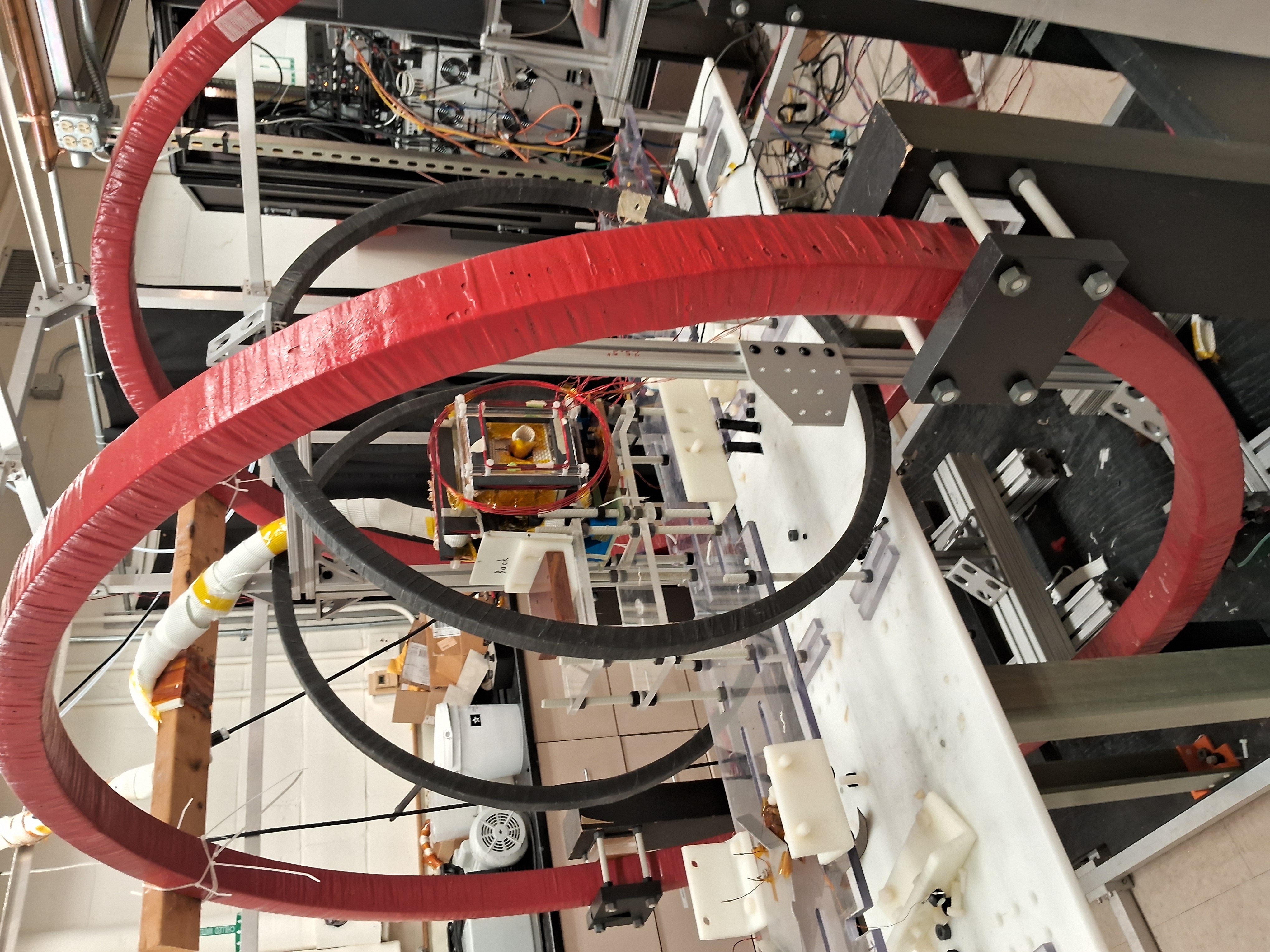} 
    \caption{Side view of magnetic field components used for in-situ NMR with the SEOP system, with the oven shown in the center. The outer red coils are the $B_0$ coils, the circular black coils are shim coils, and the square coils mounted in black frames around the oven are the dressing field coils.}
    \label{fig:b0_coils}
\end{figure}

Figure \ref{fig:b0_coils} shows a side view of the magnetic field system used for SEOP. The large red coils provide a holding field parallel to the laser beam direction which produces the Zeeman splitting in Rubidium. These coils are maintained at a constant current of $\SI{3}{\ampere}$, corresponding to a field strength of approximately $\SI{0.63}{\milli\tesla}$. These coils are also used for the B0 field for NMR experiments, during which the current is reduced to $\SI{0.91}{\ampere}$ so that the $\ce{^3He}$ Larmor frequency matches the pickup coil resonance frequency. This corresponds to a field strength of $\SI{0.19}{\milli\tesla}$.
The circular black coils shown in figure \ref{fig:b0_coils} are the shim coils, whose purpose is to cancel any DC field component parallel to the dressing field. These coils are not active during SEOP, but are set to roughly \SI{15}{\milli\ampere} during spin dressing experiments.

\subsubsection{Dressing Coil Design}
\begin{figure}
    \centering
    \includegraphics[width=0.7\linewidth, angle=-90]{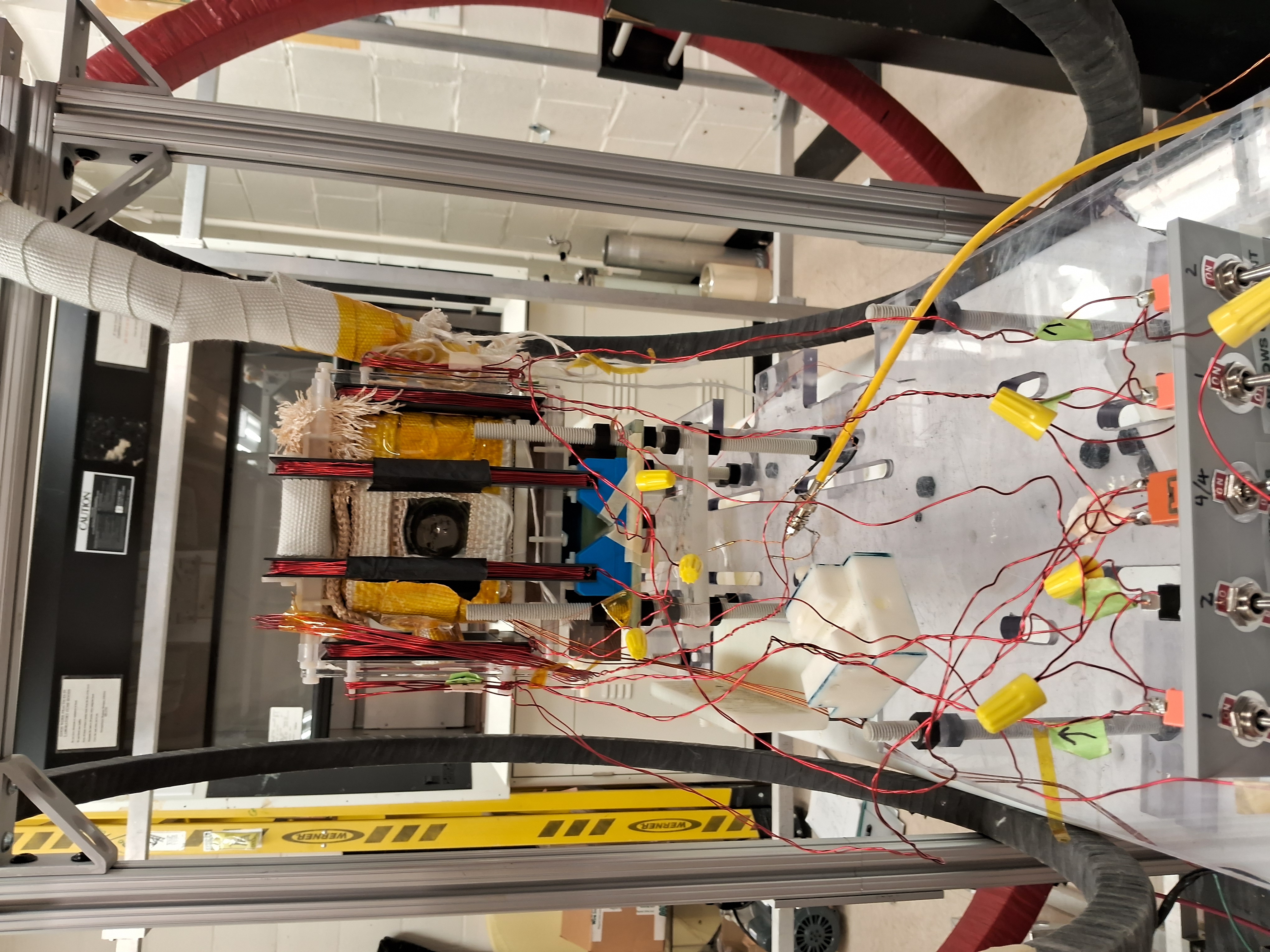}
    \caption{Front view of the SEOP oven, showing the four square dressing coils around in the oven.}
    \label{fig:oven-front}
\end{figure}

A new set of gradient and pickup coils were designed for these studies. Following the technique used by \cite{SquareCoils}, a design consisting of a set of four square coils was chosen for the dressing coils. Enforcing symmetry on the model, the parameters varied were the side lengths of the coils, the ratio of current between the inner and outer coils, and the separation between the coils. With these parameters, we can numerically solve for the configuration that will set the 2nd, 4th, and 6th order gradient of the magnetic field at the center of the coils to be zero (the odd order gradients are zero by symmetry). The optimized coil parameters are shown in table \ref{tab:coil-parameters}.
\begin{table}[]
    \centering
    \begin{tabular}{|c|c|}
        \hline
         Parameter & Value \\
         \hline
         Outer coil side length & 14.26 cm\\
         Outer coil separation & 18.25 cm \\
         Outer coil turns & 19 \\
         Inner coil side length & 21.29 cm \\
         Inner coil separation & 6.67 cm \\
         Inner coil turns & 27 \\
         \hline
    \end{tabular}
    \caption{Optimized parameters used for dressing coil design. The actual design parameters will differ from these slightly owing to the thickness of the coil wires.}
    \label{tab:coil-parameters}
\end{table}
The coils were simulated in COMSOL to confirm that the gradients were within an acceptable range, and the variation of the magnetic field along the primary field direction over the cell region was found to be less than 0.13\%. The spin dressing coil frame was 3D-printed using acrylonitrile butadiene styrene (ABS). ABS has a glass transition temperature of approximately 105 degrees Celsius \cite{ABS_Glass}, making it suitable for use around the SEOP oven. The struts which hold these ABS coils in place are machined from acrylic (PMMA). The completed and wound dressing coils are shown in figure \ref{fig:oven-front}.
The inductance of the coil was measured to be \SI{1.514}{\milli\henry} and the resistance \SI{1.39}{\ohm}. A capacitor bank with capacitance \SI{71.8}{\nano\farad} was placed in series with the coil to tune its resonance frequency to \SI{15}{\kilo\hertz}.

\subsection{SEOP and NMR Electronics}
\begin{figure}
    \centering
    \includegraphics[width=0.9\linewidth]{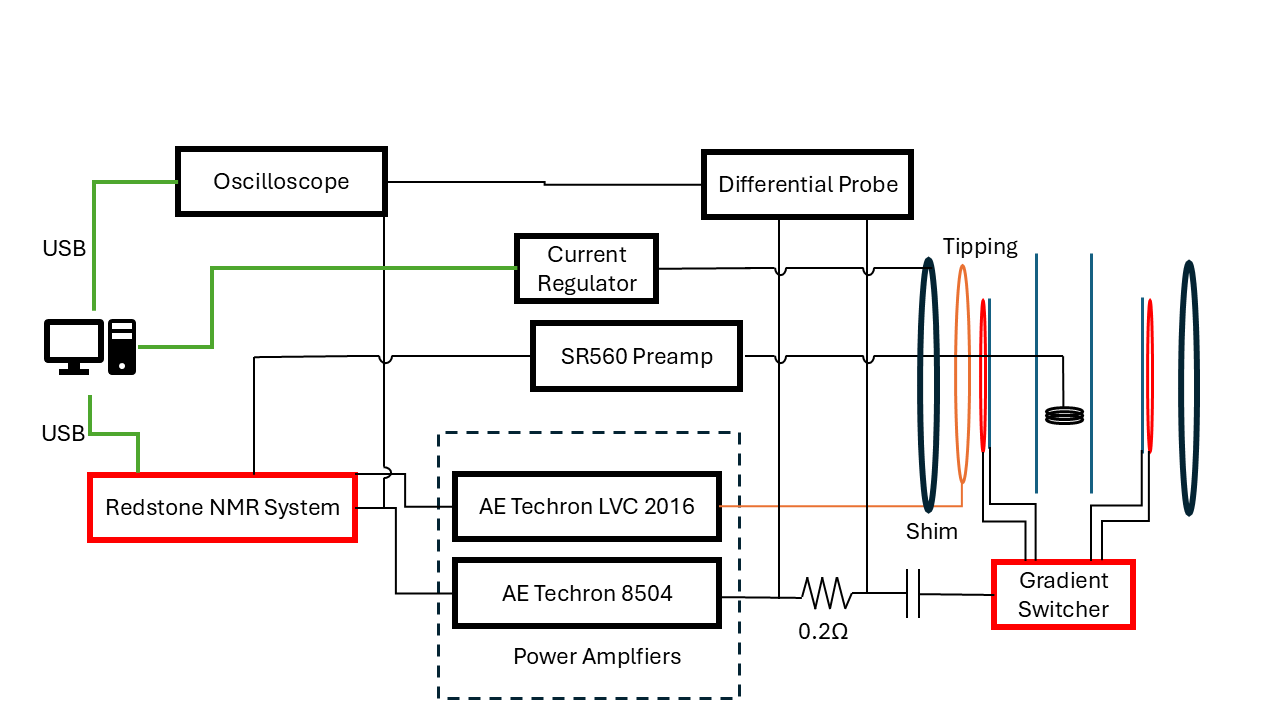}
    \caption{Simplified diagram of electronics used for NMR in the SEOP Lab}
    \label{fig:SEOP_electronics}
\end{figure}

\begin{figure}
    \centering
    \includegraphics[width=0.6\linewidth, angle=-90]{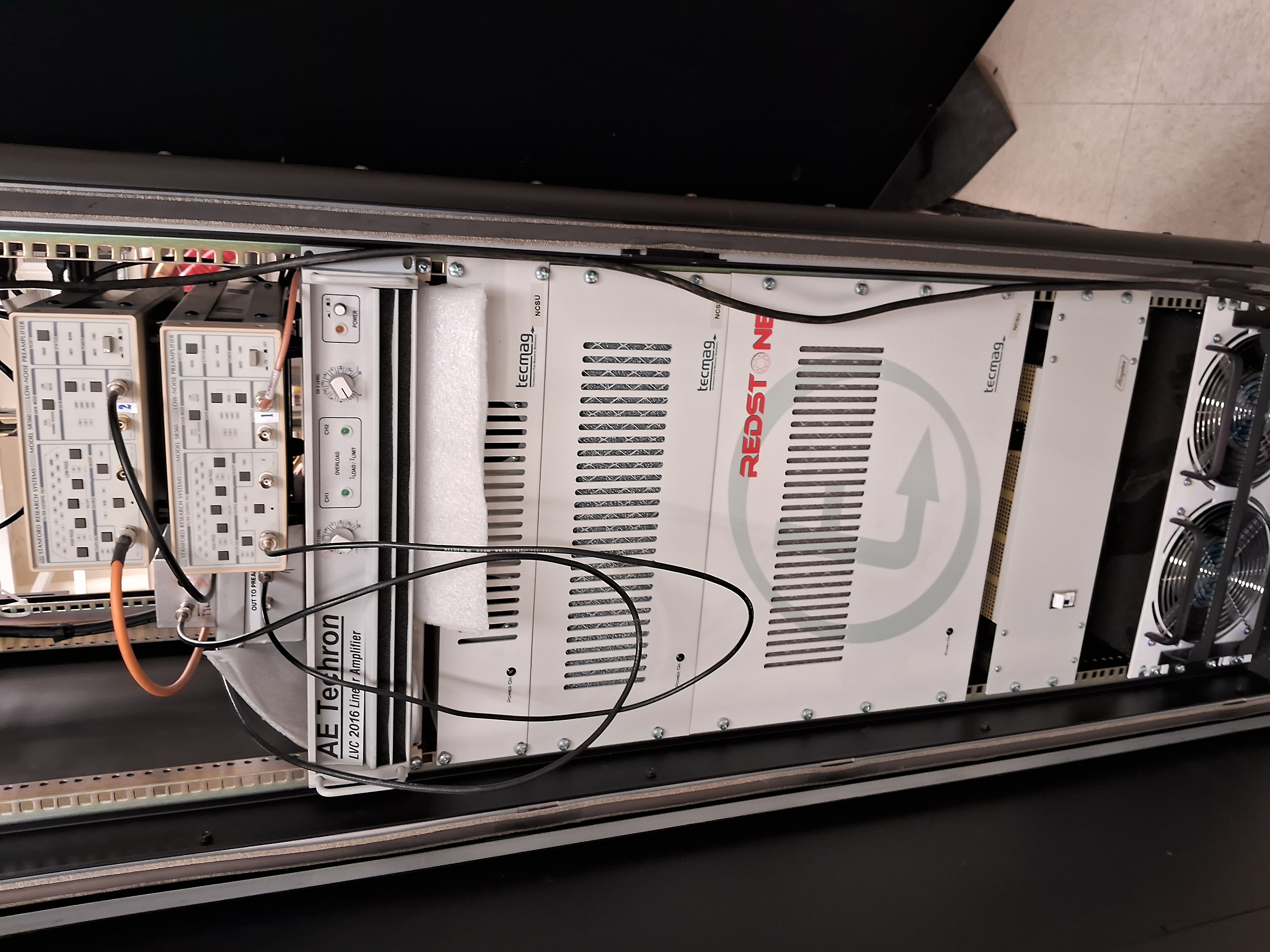}
    \caption{Electronics cabinet, including the Redstone NMR system used to perform data acquisition and arbitrary waveform generation. The LVC2016 amplifier and SR560 pre-amplifier are at the top of the cabinet.}
    \label{fig:redstone-front}
\end{figure}

A diagram of the electronics used in these measurements is shown in figure \ref{fig:SEOP_electronics}. 
% \begin{figure}
%     \centering
%     \includegraphics[width=0.5\linewidth]{figures/SEOP/TNMR_Interface.jpg}
%     \caption{TNMR software interface, showing a sample FID signal.}
%     \label{fig:tnmr-software}
% \end{figure}
Control of the dressing and tipping coils are handled by the Redstone NMR system, which is shown in figure \ref{fig:redstone-front}.
The Redstone system has many features, but for our purposes it is used as a combination arbitrary waveform generator and data acquisition module. The dressing field waveform is amplified by the AE Techron 8504 amplifier, a \SI{4}{\kilo\watt} amplifier specialized to driving reactive loads. To further improve the performance of the dressing coil system, a set of capacitors is placed in series with the dressing coil to tune the resonance frequency near \SI{15}{\kilo\hertz}. A \SI{0.2}{\ohm} resistor is also included in series with the dressing coil to monitor the current. For the tipping pulse, the reactive power requirement is much lower because the operating frequency is lower (\SI{6.2}{\kilo\hertz} vs \SI{15}{\kilo\hertz}), and because the amplitude is smaller. Thus, a LVC 2016 linear amplifier is used instead. The pickup coil signal first passes through a preamplifier, which applies a \SI{0.3}{\kilo\hertz}-\SI{30}{\kilo\hertz} bandpass filter to reduce noise.

\section{Demonstration of Magic Dressing}
In this section we describe the successful demonstration of single-species magic dressing. To do this, we artificially apply a dressing field gradient to the SEOP cell and show that a dressing field amplitude can be found that, regardless of gradient magnitude, maximizes the $T_2$ relaxation time of the system. We conclude with a discussion of the role of magic dressing for a nEDM measurement.
\subsection{Dressed Tipping}
\label{sec:dressed-tipping}
\begin{figure}
    \centering
    \includegraphics[width=0.7\linewidth]{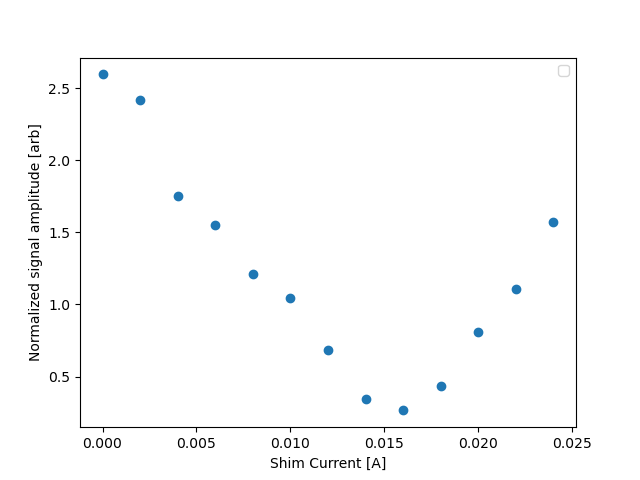}
    \caption{Sample plot demonstrating the shim coil calibration process. The shim coil current is adjusted until the size of the signal caused by the dressed tipping effect is minimized.}
    \label{fig:shim_calibration_sample}
\end{figure}

One of the most pernicious difficulties in performing this measurement was a phenomenon we term ``dressed tipping.'' This refers to the off-resonant excitation of the spins due to a dressing pulse which occurs when the dressing field and the holding field are not exactly orthogonal. This effect is most easily understood in terms of effective magnetic fields. We consider a magnetic field of the form
\begin{equation}
    B(t) = B_0 \hat{z} + \delta B\hat{x} + B_1\sin(\omega t) \hat{x}
\end{equation}
where $\delta B$ represents misalignment between the dressing and holding fields. Treating $\delta B$ as a perturbation, we find that this corresponds to the effective magnetic field
\begin{equation}
    B_{\text{eff}}(t) = J_0(\gamma B_1/\omega)B_0 \hat{z} + \delta B\hat{x}.
\end{equation}
The key point here is that as the $J_0$ term approaches zero, the effective magnetic field direction strays from $\hat{z}$. Thus, spins which are initially aligned along $\hat{z}$ will precess under the effective magnetic field and subsequently become depolarized if the dressing field is suddenly removed. It should be noted that nEDMSF operates with $\gamma_3 B_1/\omega \approx 1.3$, which is far from the first zero of $J_0$, so this effect is unlikely to be a problem for the experiment in critical dressing mode. Two techniques were employed to mitigate this phenomenon. First, dressing amplitudes that would put the system near the zero of the Bessel function were avoided; and second, a pair of shim coils were used to reduce the magnitude of $\delta B$. The shim coils - the circular black coils shown in figure \ref{fig:b0_coils} - provide a field in the $\hat{x}$ direction, whose magnitude is adjusted at the start of each measurement cycle to minimize the size of the dressed tipping effect. For these calibration runs, no tipping pulse is used, as we want to isolate the effect of the dressed tipping phenomenon. Figure \ref{fig:shim_calibration_sample} shows a sample calibration figure. The necessary shim current was found to be relatively consistent, only varying around 5 mA from day to day. The current regulator used to adjust the shim current was designed with current stability as a top priority \cite{MattMoranoThesis}. The device can be controlled using serial commands via a USB interface, which allowed for the calibration process to be automated.

\subsection{Measurement Scheme}
\label{sec:measurement-scheme}
Even with the shim coils, the dressed tipping phenomenon discussed above still manifests at larger dressing amplitudes. We thus use the following measurement sequence to investigate magic dressing. The sequence consists of four subsequences:
\begin{enumerate}
    \item Tipping Pulse, Rest, Acquire
    \item Rest, Dressing Pulse, Acquire
    \item Tipping Pulse, Rest, Acquire
    \item Tipping Pulse, Dressing Pulse, Acquire
\end{enumerate}
These four are repeated for each dressing amplitude under study. Subsequences (1) and (3) serve as normalization runs for subsequences (2) and (4) respectively, allowing us to account for the inevitable loss of polarization that occurs during the course of the experiment. Subsequence (2) is used to cancel the effect of dressed tipping; by subtracting the normalized signal acquired in step (2) from the normalized signal acquired in step (4), we can isolate the impact that the dressing pulse has on the spins that were tipped by the tipping pulse, and can eliminate the signal arising from spins that are tipped by the dressing pulse. This process is summarized in equation \ref{eq:sequences}.
\begin{equation}
    \label{eq:sequences}
    P = \frac{A_4}{A_3} - \frac{A_2}{A_1}
\end{equation}
$A_{\{1-4\}}$ denote the amplitude from subsequences 1-4 respectively, and $P$ is the polarization remaining relative to the initially tipped sample (after the tipping pulse, but before the dressing pulse).

\subsection{Analysis of Magic Dressing Data}
\begin{figure}[!h]
    \centering
    \includegraphics[width=0.7\linewidth]{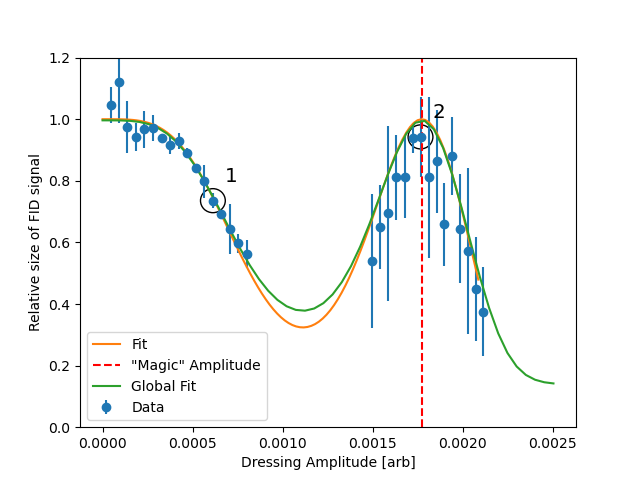}
    \caption{Polarization of the \ce{^3He} spins after a 10-millisecond dressing pulse at \SI{15}{\kilo\hertz}. All gradient coils are disconnected.}
    \label{fig:gradient2}
\end{figure}

\begin{figure}[!h]
    \centering
    \includegraphics[width=0.7\linewidth]{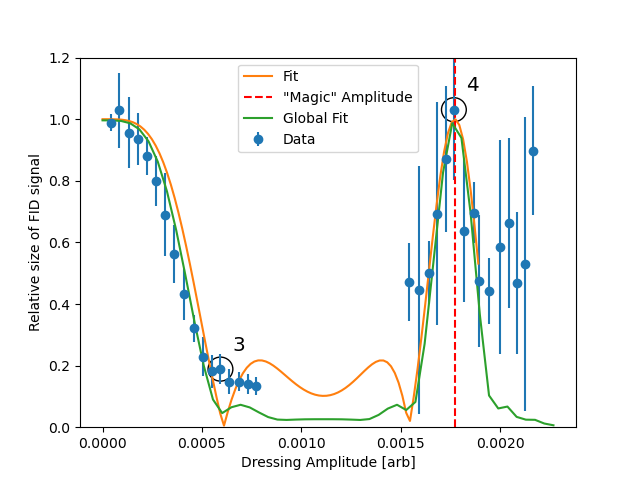}
    \caption{Polarization of the \ce{^3He} spins after a 10-millisecond dressing pulse at \SI{15}{\kilo\hertz}. Four gradient coils are active on each side of the other, each in reverse orientation.}
    \label{fig:gradient-4}
\end{figure}

Figures \ref{fig:gradient2} and \ref{fig:gradient-4} show the polarization remaining after a dressing pulse, relative to the initial polarization. The two plots differ in that for Figure \ref{fig:gradient-4}, we increase the magnitude of the AC gradient using a pair of coils in anti-Helmholtz configuration outside of the oven and in series with the dressing coil (see the red coils in figure \ref{fig:SEOP_electronics}).
For the global fit, Monte Carlo integration is used to approximate the longitudinal decay behavior of the \ce{^3He} spins. It is assumed that over the duration of the spin dressing pulse (\SI{10}{\milli\second}), the motion of the \ce{^3He} is negligible. This is justified because the diffusion constant of \ce{^3He} at room temperature is approximately \SI{1.759}{\centi\meter\squared\per\second} \cite{self_diffusion_constants}, which over \SI{10}{\milli\second} corresponds to a rms change in position by only \SI{0.13}{\centi\meter}. The interior of the cell is approximated to be a \SI{4}{\centi\meter} sphere, accounting for the thickness of the glass. We also account for the inhomogeneous response of the pickup coil. This is approximated using the fact that the sensitivity of a pickup coil to a dipole at a point in space is proportional to the strength of the magnetic field that would be produced by a current through the pickup coil at that same point in space. For a given current, we call this field $\Vec{B}_{\text{pickup}}(\Vec{r})$ and model it using Magpylib. We approximate the AC gradient near the center of the coil as being linear in position, and we assume that the static gradient is negligible, as the $T_2$ time of the SEOP system without dressing is on the order of 1 second. We use the Bessel function approximation to determine the phase shift caused by AC gradient. Putting this all together, the cell polarization is given by
\begin{multline}
    P(t) = \frac{1}{A} \biggl|\int_V dV \left(\Vec{B}_{\text{pickup}} (\Vec{r})\cdot \hat{y}\right) \\
    \times \exp\left[\frac{it\gamma^2 B_0}{\omega} J_1\left(\frac{\gamma B_1}{\omega}\right) \Vec{G_x} \cdot \Vec{r}\right]\biggr|,
\end{multline}
where
\begin{equation}
    A = \abs{\int_V dV \Vec{B}_{\text{pickup}}(\Vec{r}) \cdot \hat{y}}.
\end{equation}
The integral is taken over the SEOP cell volume $V$, and the origin ($\Vec{r} = (0, 0, 0)$) is assumed to be the center of the cell. $B_0$ and $B_1$ are the nominal holding field and dressing field amplitudes at the center of the cell, the former determined by the Larmor frequency and the latter determined based on the current measured across the shim resistor in series with the dressing coil. $\Vec{G_x}$ is the vector AC gradient $\left(\pdv{B_1}{x}, \pdv{B_1}{y}, \pdv{B_1}{z}\right)$, which is modeled as a sum of two terms: a term independent of the number of gradient coils, $\Vec{G}_\text{base}$, and a term proportional to the number of gradient coils $n$, $n\Vec{G}_\text{coil}$. $\Vec{G}_\text{base}$ is a fitted parameter, while $\Vec{G}_\text{coil}$ is modeled in Magpylib based on the known coil dimensions. The reason for this is that $\Vec{G}_\text{base}$ is intended to model imperfections in the dressing coil construction which lead to gradients; thus, its leading-order terms cannot be inferred from models of the dressing coil; meanwhile, the leading-order gradient terms for the gradient coils are due to their geometry.

For comparison, we also include a plot of a simpler $T_2$ model, where the polarization is given by
\begin{equation}
    P(t) = \abs{\operatorname{sinc}\left(\frac{it\gamma^2 B_0}{\omega \pi} J_1\left(\frac{\gamma B_1}{\omega}\right) \abs{G_x} \cdot \Delta x \right)}
\end{equation}
where $|G_x|$ is a fitted parameter representing the average gradient magnitude, $\Delta x$ is the cell diameter, and $\operatorname{sinc}(x)$ is the sinc function, defined as $\operatorname{sinc}(x) = \sin(x)/x$. For this model, rather than performing a global fit for data across different numbers of gradient coil turns, we fit to each gradient coil setting individually. This gives us a rough approximation of the gradient magnitude for each setting.

The results for transverse relaxation time are summarized in Table \ref{tab:t2}. The first two rows correspond to two points from figure \ref{fig:gradient2}, which has the smaller gradient, while the last two rows correspond to the larger AC gradient of figure \ref{fig:gradient-4}. The points without magic dressing (rows 1 and 3) were chosen to have a similar value of the dressing parameter as would be required for critical dressing in nEDMSF ($x_3 = \gamma_3B_1/\omega \approx 1.3$). The corresponding points are numbered in Figures \ref{fig:gradient2} and \ref{fig:gradient-4}. In either case, it is seen that magic dressing increases the $T_2$ time, though the effect is more pronounced when the gradient is larger.

\begin{table}[]
    \centering
    \begin{tabular}{c|c|c|c|c}
    \makecell{Plot \\ Label} & $B_1$ [\unit{\micro\tesla}] &  \makecell{Gradient \\{} [\unit{\micro\tesla\per\centi\meter}]} & \makecell{T2 [ms] \\ (1$\sigma$ range)} & Magic Dressing? \\
    \hline
    % $5.9 \times 10^2$ &               4.1 &      5 - 7 & No \\
    % $1.8\times 10^3$ &                12 &     $>43$ & Yes \\
    % $6.1 \times 10^2$ &               1.6 &     28 - 35 & No \\
    % $1.8 \times 10^3$ &               4.6 &      $>46$ & Yes \\
    1 & $6.1 \times 10^2$ & 2.1 & 28 - 35 & No\\
    2 & $1.8 \times 10^3$ & 6 & $> 46$ & Yes \\
    3 & $5.9 \times 10^2$ & 5.9 & 5 - 7 & No\\
    4 & $1.8 \times 10^3$ & 18 & $> 43$ & Yes\\

    \end{tabular}
    \caption{Estimated transverse relaxation times for selected gradient and amplitude values. The second and fourth row correspond to magic dressing, while the first and third row correspond to the nEDMSF critical dressing condition \cite{nEDMSNS}. The data points corresponding to rows in this table are numbered in Figures \ref{fig:gradient2} and \ref{fig:gradient-4}.}
    \label{tab:t2}
\end{table}

\section{Discussion of Results}
From these plots, the effects of single-species magic dressing can be seen clearly. The first feature to note is that the polarization of the sample, and thus the $T_2$ time, decreases as the dressing amplitude increases. This is consistent with the idea that that variations in the effective gyromagnetic ratio caused by spatial inhomogeneities in the dressing field amplitude - inhomogeneities that become proportionally larger as the dressing field amplitude is increased - are responsible for transverse relaxation. Second, at a particular magnitude for dressing amplitude corresponding to the first root of the Bessel function of order one ($J_1(x)$), nearly all of the polarization is recovered, regardless of how many turns of the gradient coils are applied. This, in spite of the fact that the spatial inhomogeneities should be significant at this amplitude, indicates that the magic dressing condition has been achieved.

\subsection{Magic Dressing for Two Species}
As described, magic dressing can only be used on one species at a time, as the zero of the $J_1$ Bessel function will differ for two species with different gyromagnetic ratios. Instead, one can apply a dressing field such that the difference between the derivatives of the two Bessel functions is zero. In other words, we want 
\begin{equation}
    \label{eq:magic_condition}
    \dv{\omega_n}{B_1} = \dv{\omega_3}{B_1},
\end{equation}
where $\omega_n$ and $\omega_3$ are the dressed Larmor frequencies of the neutron and \ce{^3He} respectively, given by 
\begin{equation}
    \omega_{n, 3} = \gamma_{n, 3} B_0 J_0\left(\frac{\gamma_{n, 3} B_1}{\omega}\right).
\end{equation}

\begin{figure}
    \centering
    \includegraphics[width=0.7\linewidth]{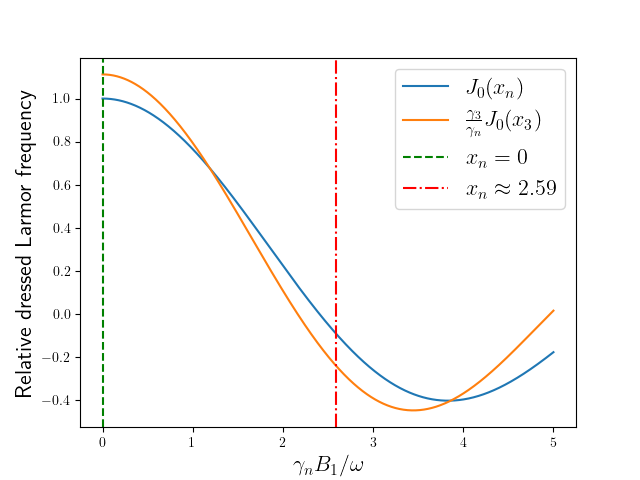}
    \caption{Relative precession frequency for neutron and \ce{^3He} as a function of dressing amplitude ($B_1$). Dressing amplitudes where the frequency shifts due to changes in dressing amplitude are equal for neutron and \ce{^3He} are shown as dashed and dotted lines.}
    \label{fig:magic_two_species}
\end{figure}

The first two solutions (including the trivial solution, $B_1 = 0$), are shown in figure \ref{fig:magic_two_species}. Neither solution is useful as a dressing scheme on its own, as neither solution corresponds to critical dressing, and so we cannot take advantage of the sensitivity gains afforded by critical dressing.

However, if we modulate between the two solutions shown in figure \ref{fig:magic_two_species}, then we can adjust the amount of time we spend at each solution to achieve both critical and magic dressing. Because the Bessel function model is only an approximate description of the dressed spin system \cite{CohenHarocheDressing1969}, it is necessary to numerically optimize a parametrized modulation function. We proceed in two steps: first, we find a pulse shape for which the magic dressing condition is achieved (ignoring the critical dressing condition). Then, we adjust the time interval between pulses in order to achieve critical dressing. Because the $B_1$ field will be off between pulses (and because $x=0$ trivially satisfies the \ref{eq:magic_condition}), this second step will not disturb the magic dressing achieved in the first step.
The modulation envelope is a smoothed square wave, given by
\begin{multline}
    B_{\text{pulse}}(t) = \frac{A}{2} \sin(\omega t + \phi)\\ 
    \times \left\{1 + \tanh\left[\frac{3\tau}{2\pi\alpha} \left(\sin\left(\frac{2\pi}{3\tau} (t - 4\alpha) + \frac{\pi}{6} \right) - \frac{1}{2}\right)\right]\right\} ,
\end{multline}
where the parameters $A$, $\phi$, $\tau$, and $\alpha$ are optimized. $A$ is the average amplitude of the $B_1$ field, $\phi$ controls the relative phase between the modulation and the carrier wave, $\tau$ controls the duration of the pulse, and $\alpha$ controls the time taken to go between the maximum and minimum amplitude. To optimize these parameters, we must define an objective function. The objective function has three terms: first, the deviation of the neutron spin vector out of the plane of precession; second, the deviation of the \ce{^3He} spin vector out of the plane of precession; and third, a term proportional to the derivative of the phase difference between the neutron and \ce{^3He} with respect to dressing amplitude. The first two terms are computed by integrating the Bloch equations and taking the $z$-component of the spin vectors at the end of the simulation time. The final term is computed by running two additional simulations, with slightly increased and slightly decreased dressing amplitude. In summary, the objective function is given by
\begin{multline}
    \label{eq:pulsefunction}
    \operatorname{Obj}(A,\phi,\tau,\alpha) = (\hat{z} \cdot \vec{\sigma}_n)^2+ (\hat{z} \cdot \vec{\sigma}_3)^2 + \\ ||U_{n+}U_{3+}^T - U_{n-}U_{3-}^T||^2,
\end{multline}
where $\vec{\sigma_n}$ and $\vec{\sigma_3}$ are the neutron and \ce{^3He} spin vectors at the end of the simulation time, and the matrices $U_{n+}$ ($U_{n-}$) and $U_{3+}$ ($U_{3-}$) are the propagators for the neutron and \ce{^3He}, simulated with a 1\% larger (smaller) value for $A$. The operator $||\cdot||^2$ is the Frobenius norm \cite{MatrixComputations}. The optimization was conducted using the SAMIN() simulated annealing algorithm available in the Optim.jl package in Julia \cite{Optim.jl}, and the Bloch equations were integrated with a custom-made Julia package. The optimized parameter values are given in table \ref{tab:Magic Parameters}.

\begin{table}
\caption{Optimized magic dressing parameters for neutron and \ce{^3He}, assuming a dressing frequency of $\omega = \qty{1}{\kilo\hertz}$ and a holding field strength of $B_0 = \qty{5}{\micro\tesla}$}
\label{tab:Magic Parameters}
\centering
\begin{tabular}{c|c|c}
     Parameter & Value & Units \\
     \hline
     $A$ & 0.11176 & \unit{\milli\tesla} \\
     $\phi$ & 1.3941 & \\
     $\tau$ & 7.3122 & ms\\
     $\alpha$ &  1.4754 & ms\\
     $t_{\text{rest}}$ & 11.149 & ms\\
\end{tabular}
\end{table}

\begin{figure}
    \centering
    \includegraphics[width=0.7\linewidth]{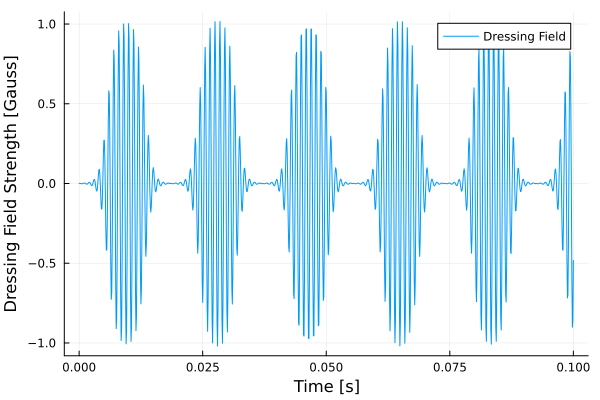}
    \caption{Two-species modulated magic dressing pulse with optimized parameters.}
    \label{fig:magic_pulse_train}
\end{figure}

\begin{figure}
    \centering
    \includegraphics[width=0.7\linewidth]{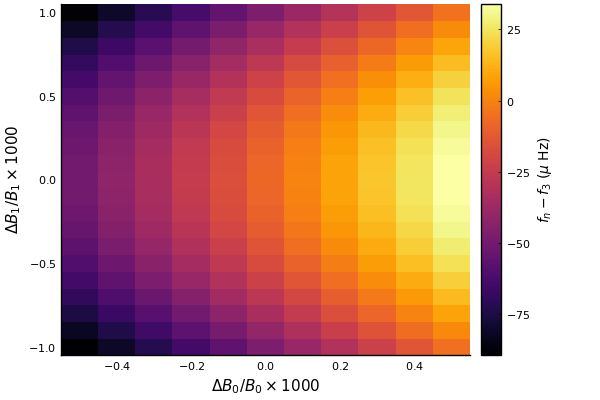}
    \caption{Heatmap of the frequency difference between neutron and \ce{^3He} as holding field strength and dressing amplitude are varied about the magic dressing parameters.}
    \label{fig:perturbation_heatmap}
\end{figure}

With the pulse shape optimized, we next optimize the time between pulses $t_{\text{rest}}$, using binary search to ensure critical dressing. This is given by the parameter $t_{\text{rest}}$. The final, periodic pulse sequence then has the form
\begin{equation}
    B_x(t) = B_{\text{pulse}}(t ~\%~(t_{\text{rest}}  + \tau))
\end{equation}
where $a~\%~b$ means to take the remainder of $a$ when divided by $b$. The magnetic field strength as a function of time is shown for \qty{100}{\milli\second} in figure \ref{fig:magic_pulse_train}.
To confirm that this pulse train reduces depolarization due to magnetic field amplitude fluctuations, we simulate neutrons and \ce{^3He} spin vectors over time for various perturbations in $B_0$ and $B_1$ about the magic dressing parameters above. A heatmap of the frequency difference for these perturbations is shown in figure \ref{fig:perturbation_heatmap}. It is worth noting that near the nominal magic dressing parameters (the center of the heatmap), the variation with respect to $B_1$ is quadratic, indicating that magic dressing has been achieved.

Two-species magic dressing is suitable for an nEDM measurement when a comagnetometer species is present, as is the case for nEDMSF \cite{nEDMSNS}. This allows one to reap the benefits of critical dressing, while loosening the amplitude stability requirement imposed on the dressing field. This can make it easier to procure a power supply that can provide the necessary current. 

\section{Conclusion}
In this work we have developed a method of spin dressing which can be employed to extend the transverse relaxation time of a single spin species in the presence of AC magnetic field gradients. We demonstrated this technique by applying an AC gradient to a sample of polarized \He{}. The sample was produced using a spin-exchange optical pumping apparatus with custom-designed in situ NMR coils. When magic dressing was utilized, it was found that nearly all of the initially tipped population was still polarized at the end of the dressing pulse. Finally, we considered how these principles could be generalized to the simultaneous magic and critical dressing of two spin species as would be relevant for neutron electric dipole moment (nEDM) experiments such as nEDMSF, and developed a dressing modulation scheme that could achieve both of these goals.

\section{Acknowledgments}
The author thanks Haiyan Gao, for generously providing access to the Duke SEOP lab, and Zhiwen Zhao, whose technical expertise contributed significantly to the smooth operation of the SEOP apparatus. The author would also like to thank Brad Filippone and Robert Golub for their helpful comments and review of the manuscript. This work was funded by the National Science Foundation (NSF) Grants No. 2110898 and 1822515. 

\bibliography{references}
\end{document}